\documentclass[
	paper=a4,
	DIV=27,+-++++
	fontsize=11pt	
	]{scrartcl}

	\KOMAoption{captions}{tableheading, figuresignature, nooneline, bottombeside, outerbeside}
	\addtokomafont{caption}{\small}			
	\addtokomafont{captionlabel}{\bfseries} 
	
		 \addtokomafont{author}{\small}
		 \addtokomafont{date}{\small}
		 \addtokomafont{publishers}{\small}

\usepackage[utf8]{inputenc} 
\usepackage[T1]{fontenc}
\usepackage[english]{babel}
\usepackage{ulem} 
\usepackage[sfdefault, light, condensed]{roboto}	    
\usepackage{amsmath}				
\usepackage{amssymb}				
\usepackage{siunitx}	            
	\DeclareSIUnit{\angstrom}{\textup{\AA}}
\usepackage[version=4]{mhchem}      
\usepackage[onehalfspacing]{setspace} 
\usepackage{graphicx}
	\graphicspath{ {images/} }
\usepackage{authblk}	
\usepackage{textgreek}	
    \usepackage{upgreek}
\usepackage[dvipsnames]{xcolor}
\usepackage{hyperref}

\usepackage[sort&compress,square,numbers,comma]{natbib}

\title{Solving the Phase Problem of Diffraction: \\ X-ray Standing Waves Imaging on Bismuthene/SiC(0001)}

    \author[1,2]{Niclas Tilgner}
    \author[1,2]{Susanne Wolff}
    \author[3,4]{Serguei Soubatch}
    \author[5]{Tien-Lin Lee}
    \author[1,2]{Fabian G\"{o}hler}
    \author[3,4,6]{F.\ Stefan Tautz}
    \author[1,2]{Thomas Seyller}
    \author[1,2,*]{Philip Sch\"{a}dlich}
    \author[3,4,6,$\dagger$]{Christian Kumpf}
	
    \affil[1]{Institute of Physics, Chemnitz University of Technology, 09126 Chemnitz, Germany}
    \affil[2]{Center for Materials, Architectures and Integration of Nanomembranes (MAIN), 09126 Chemnitz, Germany}
    \affil[3]{Peter Gr\"{u}nberg Institut (PGI-3), Forschungszentrum J\"{u}lich, 52425 J\"{u}lich, Germany}
    \affil[4]{J\"{u}lich Aachen Research Alliance (JARA), Fundamentals of Future Information
Technology, 52425 J\"{u}lich, Germany}
    \affil[5]{Diamond Light Source Ltd., Harwell Science and Innovation Campus, Didcot, Oxfordshire, OX11 0DE, United Kingdom}
    \affil[6]{Experimentalphysik IV A, RWTH Aachen University, 52074 Aachen, Germany}
    \affil[*]{\textit{Corresponding author E-Mail:} philip.schädlich@physik.tu-chemnitz.de}
    \affil[$\dagger$]{\textit{Corresponding author E-Mail:} c.kumpf@fz-juelich.de}

\begin{document}
	
	\maketitle
	
	\medskip
	\textbf{Keywords:} 
	\emph{X-ray standing wave imaging, phase problem of diffraction techniques, bismuthene, graphene} \par
	
	\begin{abstract}
     The phase retrieval problem is a fundamental shortcoming of all diffraction-based methods, arising from the inability to measure the phase of scattered waves.
     The (normal incidence) X-ray standing wave (NIXSW) technique circumvents this issue by introducing a (Bragg-generated) X-ray standing wave field throughout the sample, relative to which any atomic species can be localized by probing its fluorescence or photoelectron yield. In essence, in a single measurement the complex scattering factor (i.e., its amplitude \textit{and} phase) corresponding to the used Bragg reflection is determined. Performing this for multiple Bragg reflections enables one to reconstruct the scattering density of the sample in three dimensions, straightforwardly as the Fourier sum of all measured (complex) scattering factors. 
     Here, we utilize this technique to reveal the structural key features involved in the formation of the quantum spin Hall insulator bismuthene on silicon carbide. In this prominent example, the two-dimensional Bi layer is confined between a 4H-SiC substrate crystal and an epitaxial graphene layer. The key finding is a change in the adsorption site of the Bi atoms underneath the graphene upon hydrogenation, caused by the H-saturation of one (out of three) Si dangling bonds per unit cell. This structural change, clearly revealed by our NIXSW imaging experiment, is the key feature leading to the formation of the characteristic band structure of the 2D bismuthene honeycomb. 
	\end{abstract}

\section{Introduction}
    
    Resolving the structure of surfaces, interfaces and thin films at the atomic scale is a fundamental prerequisite for understanding the properties and functionalities of almost any physical system or device.  Prominent examples in this context are stacks of two-dimensional materials and van der Waals heterostructures, in particular novel materials with emerging properties, which are in the focus of intense research since the discovery of graphene about two decades ago. 
    For the detailed structure determination in this (and many other) fields, various diffraction methods play a crucial role. However, it is well known that any diffraction experiment suffers from the fact that it can only determine the amplitudes of the scattering factors, since the measured intensities represent the absolute squares of the complex scattering factors, while the phases are lost. If amplitudes and phases of the structure factors were available, a simple and straightforward Fourier-backtransformation would be possible, resolving the atomic structure of the system in unrivaled clearness, without the necessity of any structural model refinement, and hence without the remaining uncertainty of whether or not the unique ``best structure'' has been found in the refinement.
    
    While in the past significant effort was invested in developing suitable workarounds for the phase problem, such as ab initio approaches (so called ``direct methods'') \cite{arnal2019ab, fienup1982phase, elser2003phase, oszlanyi2004ab} or the Patterson method \cite{patterson1934fourier}, there is one very elegant way of determining amplitudes and phases of the scattered waves in crystalline samples, namely the (Normal Incidence) X-ray Standing Waves ((NI)XSW) technique \cite{batterman1969detection, zegenhagen1993surface, vartanyants2001theory, Woodruff_2005, Zegenhagen2013}.
    This technique exploits the interference pattern created by incident and Bragg reflected X-rays, which represents a standing wave field with a period corresponding to the Bragg plane spacing of the used reflection. The atoms within the structure can then be localized with respect to this standing wave field, since they absorb differently strong from the standing wave depending on their positions. The absorption yield is detected via photoelectrons or fluorescence. 

    The NIXSW method, as described so far, allows the localization of the atomic species in one dimension, namely perpendicular to the Bragg planes used to generate the standing wave, that is the direction of the scattering vector $\boldsymbol{H}=$ ($hkl$).  In most of the recent NIXSW based studies on atomic and molecular adsorbates at surfaces and on 2D material heterostacks \cite{LinPhys.Rev.B2022, SchaedlichAdv.Mater.Interfaces2023, Wolff_2024, grossmann2022evolution, ding2023does}, a reflection normal to the surface was used, in order to investigate vertical distances between adsorbates or layers and the substrate. These vertical distances were then interpreted in terms of bonding distances, revealing valuable insights into the nature of chemical interactions at surfaces (e.g., covalent or van der Waals). In some favorable cases it was possible to obtain also some lateral information, by utilizing reflections with a scattering vector not normal to the surface. This allows, e.g., to determine adsorption sites through triangulation \cite{hBN_on_Ni, XSW_of_Si_on_GaAs, woicik1994structural}. However, this technique is often not feasible, in particular for systems with different surface terminations or multiple adsorption sites.

    Nonetheless, the fundamental brilliance of this approach is the fact that every single of these ``one dimensional'' NIXSW measurements on the scattering vector $\boldsymbol{H}$ in fact provides two parameters, namely amplitude and phase of the $\boldsymbol{H}^{\text{th}}$ (complex) structure factor $\mathcal{F}_{\boldsymbol{H}}$ of that atomic species, the absorption yield of which was recorded. Since the structure factor $\mathcal{F}_{\boldsymbol{H}}$ in turn represents the $\boldsymbol{H}^{\text{th}}$ Fourier component of the three-dimensional atomic distribution \cite{bedzyk2004x, Woodruff_2005}, a sufficiently complete set of individual NIXSW measurements on different (inequivalent) Bragg reflections $\boldsymbol{H}$ opens the way to a Fourier analysis of the data. Hence, the atomic density in the crystal can be reconstructed, simply by calculating the Fourier sum of all experimentally available structure factors.  This Fourier-based reconstruction is called ``NIXSW imaging'', since it represents a model-free approach yielding a three-dimensional image of the atomic arrangement within the unit cell with sub-Angstrom resolution \cite{Bedzyk_phase_structure_factor, bedzyk2004x, Woodruff_2005}. NIXSW imaging is capable of revealing intricate structural details in cases when conventional diffraction or microscopy methods may struggle to yield sufficient information, as we will demonstrate for the prominent example of bismuthene on SiC in the following. Note that, although this technique was suggested more than two decades ago \cite{bedzyk2004x}, it was hardly used owing to the fact that in-vacuum high-precision goniometers for reliable and reproducible sample orientation are required, but were not available until recently.

    In this paper, we report an NIXSW imaging experiment on a 2D heterostack consisting of a graphene layer on an intercalated Bi layer on 4H-SiC(0001), a system which at present attracts highest interest in the 2D materials community since it is discussed to form the 2D material bismuthene, a potential quantum spin Hall insulator. And in fact, the structural results we obtain in our study allowed us to identify the structural key feature that turns the intercalated Bi layer into 2D bismuthene.

    Contemporary research on confinement heteroepitaxy of Bi beneath epitaxial graphene on SiC(0001) identified two different Bi phases that are formed by intercalation of the graphene buffer layer \cite{SohnJ.KoreanPhys.Soc.2021, Wolff_2024} (which is often entitled zeroth layer graphene (ZLG) since it is hybridized with the substrate \cite{RiedlPhys.Rev.Lett.2009, BriggsNat.Mater.2020, EmtsevNat.Mater.2009}). In the intercalation process performed by Bi deposition on the sample surface and sequential annealing at elevated temperatures, at first the Bi \textalpha\ phase is formed with monolayer coverage and, in relation to the substrate's surface unit cell, ($1 \times 1$) periodicity \cite{SohnJ.KoreanPhys.Soc.2021, Wolff_2024}. At higher temperatures, the Bi density decreases and the ($\sqrt{3} \times \sqrt{3}$)$R30^{\circ}$ reconstructed \textbeta\ phase is formed \cite{SohnJ.KoreanPhys.Soc.2021, Wolff_2024}. 

    The latter is the topic of the research presented here. So far, the atomic configuration of the \textbeta\ phase has not yet been determined. 
    The initial suggestion by Sohn et al.\ \cite{SohnJ.KoreanPhys.Soc.2021} was a structure with only one Bi atom per ($\sqrt{3} \times \sqrt{3}$) unit cell (1/3 monolayer coverage), adsorbed on $\text{H}_\text{3}$ hollow sites.      
    But one might also expect a structure similar to that obtained by Reis et al.\ \cite{ReisScience2017} who reported the formation of a Bi honeycomb on a hydrogen terminated SiC surface, which the authors discuss in terms of a bismuthene layer.  
    Our own work \cite{Tilgner2025} shows that hydrogen indeed plays an important role in the formation of bismuthene. While the \textbeta\ phase, as prepared from the \textalpha\ phase, shows no electronic features that would be indicative for 2D bismuthene (as seen in ARPES), the situation changes when the sample is annealed in hydrogen atmosphere. In this case some of the surface Si dangling bonds, which before hydrogenation can only be saturated by Bi, now become saturated by hydrogen, which changes the electron configuration in the Bi \textbeta\ phase layer and triggers the formation of a honeycomb structure that shows all characteristic features of the quantum spin hall insulator bismuthene.  The decisive change in the structure of the Bi layer is a change of the Bi adsorption site from a $\text{T}_\text{4}$ hollow to a $\text{T}_\text{1}$ top site above the Si atoms, which is detected in our NIXSW imaging study and discussed in detail in this paper.

\section{Results \& Discussion}
    
  \subsection{Normal Incidence X-ray Standing Wave}

\begin{figure}[t!]
        \centering
        \includegraphics[scale=1]{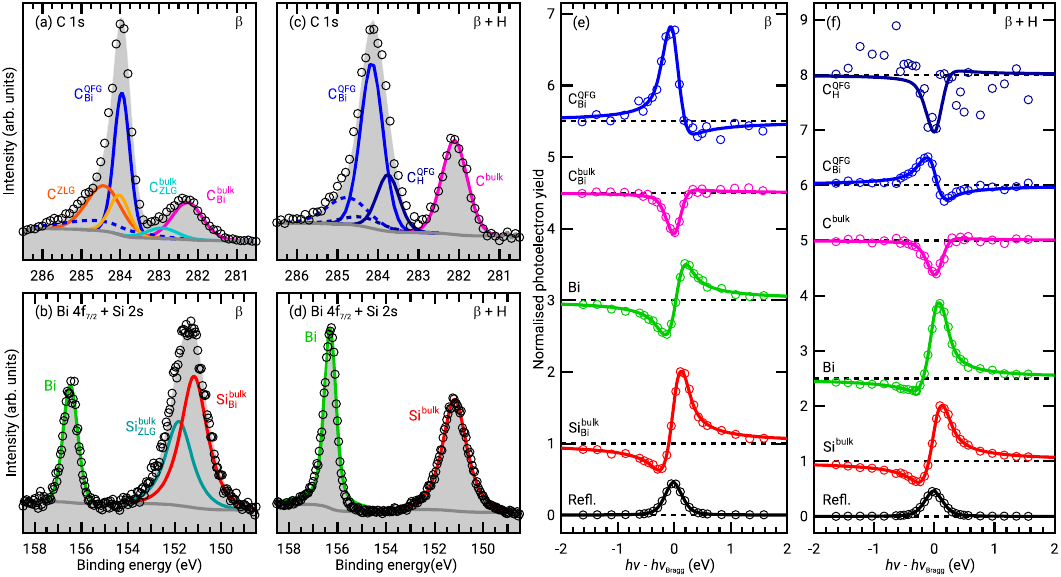}
        \caption{(a,b) and (c,d) Representative XP spectra of C\,1s, Si\,2s and Bi\,4f$_\text{7/2}$ of the \textbeta\ phase before, and the bismuthene phase after hydrogenation (\textbeta +H), respectively. The spectra were measured at a photon energy $\sim$ \qty{5}{eV} below the the Bragg energy of the 4H-SiC(0004) reflection ($h\nu_{\text{Bragg}} = \text{\qty{2.4634}{keV}}$). All peaks relevant for the NIXSW analysis are labeled, and further details on the fitting models discussed in the Supplementary Information. (e,f) Corresponding absorption yield curves and typical reflectivity curves of the (0004) reflection for both phases.   } 
        \label{fig:xps_yield_0004}
    \end{figure}

    The NIXSW technique is element specific, since the absorption yield can be measured separately for each atomic species (either by fluorescence or X-ray photoelectron spectroscopy (XPS)). Owing to a relatively high energy resolution of modern synchrotron photoemission beamlines one can even separate identical species in different chemical environments. This is essential in our case because we need to separate bulk carbon from graphene carbon. However, a very careful and detailed fitting of the XPS data is required for a proper separation of the individual species. In Figures \ref{fig:xps_yield_0004}(a-d) we show representative XP spectra for the three relevant atomic species, namely C\,1s, Si\,2s and Bi\,4f$_\text{7/2}$ core levels. Note that the latter two were measured in a single spectrum, due to their similar binding energies. Panels (a) and (b) show the data for the \textbeta\ phase (``\textbeta''), panels (c) and (d) for bismuthene (hydrogenated \textbeta\ phase, ``\textbeta+H''). For both phases, different fitting models have been used, according to different configurations of the samples. 

    When preparing the Bi \textbeta\ phase from an \textalpha-phase sample, a partial de-intercalation cannot be avoided \cite{Wolff_2024}. This leads to the coexistence of the desired Bi-intercalated quasi-freestanding graphene (QFG) domains with non-intercalated areas, in which only a graphene buffer layer (ZLG) is present. These ZLG regions are visible in both C\,1s and Si\,2s spectra. In the C\,1s spectra (\autoref{fig:xps_yield_0004}(a)), two side peaks C$^\text{ZLG}$ (red and orange) appear next to the main graphene peak C$^\text{QFG}_{\text{Bi}}$ (blue), and also next to the C\,1s-bulk peak (C$^\text{bulk}_{\text{Bi}}$, magenta) a side peak C$^\text{bulk}_{\text{ZLG}}$ (cyan) is seen. Also in the Si\,2s spectrum the core level splitting caused by the different structures ZLG and QFG above can be seen (red and dark green curves in \autoref{fig:xps_yield_0004}(b)) \cite{EmtsevNat.Mater.2009}. For fitting the Bi\,4f$_\text{7/2}$ core level, a single symmetric profile was used, representing the confined insulating Bi layer of the \textbeta\ phase \cite{Wolff_2024, SohnJ.KoreanPhys.Soc.2021}. Further details of these fitting models are discussed in Section 1 of the Supplementary Information. 

    The hydrogenation performed to transform the Bi \textbeta\ phase into bismuthene also affects the ZLG regions of the sample. The ZLG in these areas undergo hydrogen intercalation \cite{RiedlPhys.Rev.Lett.2009}, which transforms them into H-intercalated QFG, and causes significant changes in both the C\,1s and the Si\,2s spectra. The C$^\text{ZLG}$ component is replaced by a C$^\text{QFG}_{\text{H}}$ peak, now at a lower binding energy than that of the C$^\text{QFG}_{\text{Bi}}$ peak, and the bulk carbon appears as one single peak since the difference in the surface band bending between bulk underneath QFG and bismuthene cannot be resolved. Furthermore, it is noteworthy that in contrast to the \textbeta\ phase before hydrogenation, the Bi 4f$_\text{7/2}$ peak is now asymmetric, reflecting the metallic nature of the Bi layer after hydrogenation \cite{Tilgner2025}.
    
    In an NIXSW scan, the photon energy is varied in a range of a few eV around the Bragg energy of the applied reflection, in this case the 4H-SiC(0004). In addition to the XP spectra, the Bragg reflected intensity (so-called ``reflectivity'') is also recorded at each energy step during the scan, and typical profiles are plotted as the bottom curve in \autoref{fig:xps_yield_0004}(e) and (f). In the photon energy range with non-zero reflectivity, a standing wave field is formed by the interference of the incident and the Bragg reflected wave. During the scan through the Bragg condition, the phase of this XSW field changes by \textpi, causing the standing wave to shift by half of the Bragg plane spacing through the crystal. This in turn causes a modulation of the X-ray intensity at the position of any atom in the crystal (and at its surface), which manifests itself as a modulation of the photoelectron yield. 
    Consequently, recording XP spectra near the Bragg energy and plotting the partial photoelectron yields (for all relevant components) as a function of the photon energy provides photoelectron yield curves that are characteristic for the positions of the corresponding atoms relative to the Bragg planes \cite{Woodruff_2005}. In \autoref{fig:xps_yield_0004}(e) and (f) we display the yield curves for the two phases under study.
    
    In a conventional NIXSW data analysis, the yield profiles are fitted, together with the reflectivity curve, and two structural parameters are obtained, the so-called coherent position $P_\textrm{c}^{\boldsymbol{H}}$ and the coherent fraction $F_\textrm{c}^{\boldsymbol{H}}$, both taking values between 0 and 1. Usually, the former represents the (averaged) position of the considered species, w.r.t.\ the $\boldsymbol{H}=$ ($hkl$) Bragg planes and in units of the Bragg plane spacing $d_{\boldsymbol{H}}$, and the latter indicates the quality of order of the species. $F_\textrm{c}^{\boldsymbol{H}}=$ 1 indicates the case of perfect order (all relevant atoms at the same distance to the Bragg planes), while a value of 0 usually corresponds to complete disorder. In cases when a species obtains several well-defined positions, the situation is more complicated, as will be discussed below. 

\begin{figure}[b!]
        \centering
        \includegraphics[scale=1]{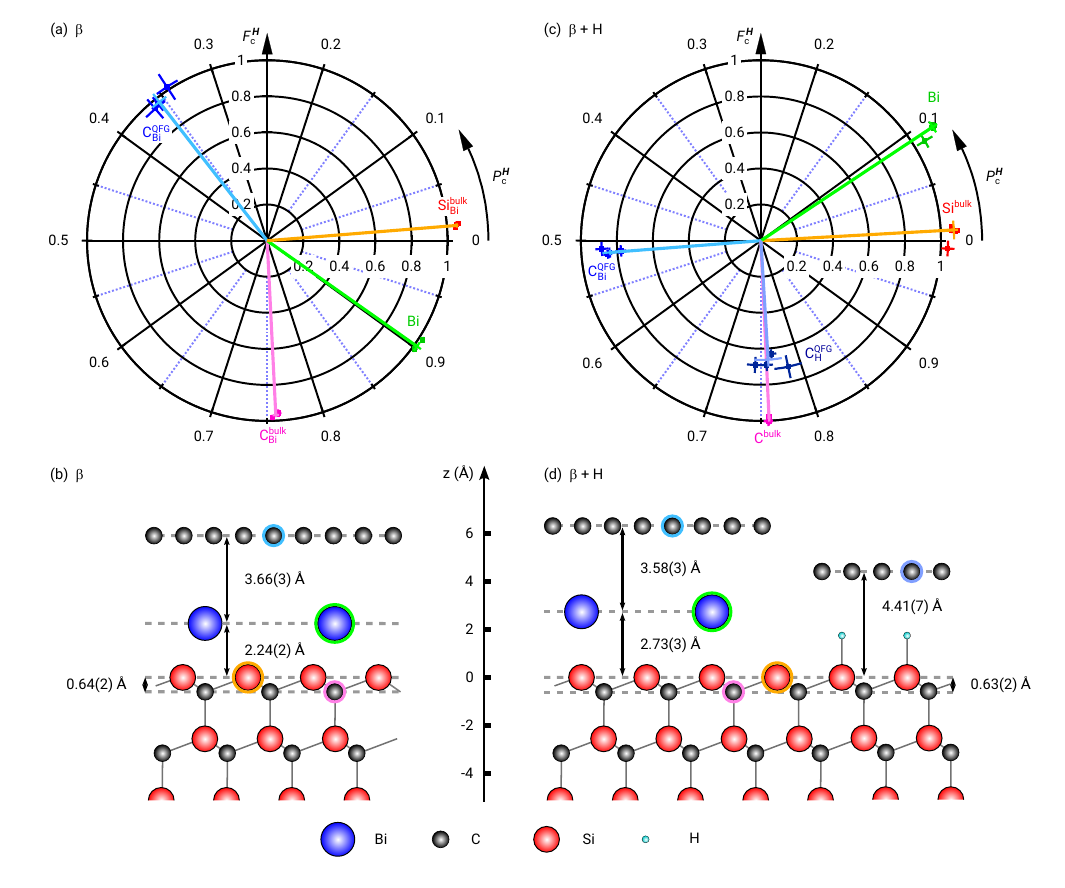}
        \caption{(a) Argand representation of the NIXSW results for the \textbeta\ phase (before hydrogenation), obtained for the (0004) reflection. Each data point represents a complex number with modulus $F_\textrm{c}^{\boldsymbol{H}}$ and phase $P_\textrm{c}^{\boldsymbol{H}}$. Individual data points for each species correspond to several measurements on different sample positions, their average is shown as polar vectors with error bars. (b) Corresponding ball-and-stick model illustrating the vertical distances between the relevant species, derived from the coherent positions. (c,d) Same like (a,b), but for the bismuthene phase after hydrogenation. In (d), bismuthene and QFG regions are illustrated in the left and right, respectively.  }
        \label{fig:argand_ball-and-stick_0004}
    \end{figure}

    It should be mentioned that the fitting parameters $P_\textrm{c}^{\boldsymbol{H}}$ and $F_\textrm{c}^{\boldsymbol{H}}$ are obtained without any modeling, but are a direct result of the experimental data.  For interpretation and illustration they are often displayed as complex numbers $F_\textrm{c}^{\boldsymbol{H}} \exp\left(2\pi i P_\textrm{c}^{\boldsymbol{H}} \right)$ in a polar diagram, the so-called Argand diagram. 
    \autoref{fig:argand_ball-and-stick_0004}(a) shows this diagram for the NIXSW results of the \textbeta\ phase obtained from the experiment on the (004) Bragg reflection ((0004) in the notation for hexagonal lattices, ($hkil$) with $i=-h-k$). 
    Each data point represents one individual measurement that has been repeated on several locations on the sample surface. Hence, the small scattering of the data for each species indicates a high sample homogeneity.

    For all four species (Si and C in the bulk, Bi and graphene-C at the surface) we find coherent fractions very close to unity. While for the bulk species this is to be expected, for both surface species, Bi and graphene, it indicates the adsorption of the atoms at a well defined height, i.e., in a very flat and unbuckled layer. Hence, the coherent positions can straightforwardly be interpreted as distances between the atomic layers, as illustrated in the ball-and-stick model of the \textbeta\ phase shown in \autoref{fig:argand_ball-and-stick_0004}(b). Most interestingly, we find a vertical spacing of only \qty{2.24(2)}{\angstrom} between the bulk-terminating Si atoms and the Bi layer, a number significantly smaller than the sum of the covalent radii of the two species (\qty{2.67}{\angstrom} \cite{pyykko2009molecular}). This distance is not compatible with a vertical covalent Bi-Si bond, making the on-top adsorption of Bi above the uppermost Si very unlikely.
    It might be compatible with a hollow site adsorption, as will be discussed below. 
    As a second result the spacing between the Bi layer and graphene was determined to be \qty{3.66(3)}{\angstrom}, a value that is in good agreement with the sum of the corresponding van der Waals radii (\qty{3.77}{\angstrom} \cite{bondi1964van, mantina2009consistent} ), and hence indicates a primarily van der Waals-like interaction between these two layers. This is consistent with previous NIXSW studies on Bi-intercalated graphene and other similar systems \cite{Wolff_2024, LinPhys.Rev.B2022, SchaedlichAdv.Mater.Interfaces2023}.

    For the bismuthene phase, the Argand diagram is shown in \autoref{fig:argand_ball-and-stick_0004}(c). The coherent positions of the surface species discussed for the \textbeta\ phase have changed significantly in the hydrogenation process; both became larger by almost the same amount, indicating that the Bi-graphene bilayer moves upwards almost rigidly. In the left part of \autoref{fig:argand_ball-and-stick_0004}(d) the numbers are shown. Bi moved up by \qty{0.49}{\angstrom} to \qty{2.73(3)}{\angstrom}, a number very close to the covalent bonding distance, and graphene is now only slightly closer to Bi, at a distance of \qty{3.58(3)}{\angstrom}. 

    Furthermore, one additional graphene species has been identified for the bismuthene sample, originating from the regions that were originally covered by a ZLG only, and have transformed into H-intercalated (QFG) regions (see discussion above). 
    Although the data points for the QFG scatter a bit more than for all other species, owing to smaller XPS intensities and more noisy absorption yield profiles, the data can be reliably analyzed and result in a height of \qty{4.41(7)}{\angstrom} and a slightly smaller coherent fraction of \qty{0.66(4)}{}. Both values closely match those obtained in previous NIXSW studies on hydrogen-intercalated graphene on SiC(0001) ($z =$ \qty{4.27}{\angstrom} and $F_\textrm{c}^{\boldsymbol{H}} =$ 0.68 \cite{SforziniPhys.Rev.Lett.2015, SforziniPhys.Rev.Lett.2016}). The corresponding ball-and-stick model is shown in the right part of \autoref{fig:argand_ball-and-stick_0004}(d). 

    The main conclusion of a comparison of the (0004) NIXSW results for the \textbeta\ phase (before hydrogenation) and the bismuthene phase (after hydrogenation) is the increase of the vertical distance between Bi and the uppermost bulk Si atoms from \qty{2.24(2)}{\angstrom} to \qty{2.73(3)}{\angstrom}. This finding is compatible with a change of the Bi adsorption site from a hollow to the on-top site, but from vertical distances alone an unambiguous conclusion is hardly possible. There are too many other factors, as e.g., a possible change in the bonding strength, that may influence the vertical distance. 
    An unambiguous determination of the adsorption site requires additional horizontal information, as it is provided by NIXSW performed on inclined reflections. In the following section we report such measurements, analyzed using the Fourier-based reconstruction of atomic densities called NIXSW imaging, which indeed results in an unambiguous adsorption site determination of Bi in both phases under study.

  \subsection{Normal Incidence X-ray Standing Wave Imaging}

    As demonstrated above for the (0004) reflection, NIXSW enables the precise determination of atomic positions relative to the Bragg planes with chemical selectivity, since separate measurements for each atomic species can be performed. In case of a single adsorption geometry this is straightforward, since the measured coherent position is directly related to the vertical position (adsorption height) of the atoms. But even in more complex situations, i.e., when multiple adsorption sites are present and this straightforward interpretation does not apply any more, NIXSW still allows a clear structure determination in many cases. The reason for this is that the parameters obtained from a single NIXSW measurement (using the Bragg reflection $\boldsymbol{H}$), that are the coherent fraction and position, are directly related to the structure factor $\mathcal{F}_{\boldsymbol{H}}$ of the structure under study. Strictly speaking, the Argand representation of the NIXSW results corresponds to the normalized species-specific geometrical structure factor $\mathcal{F}_{\boldsymbol{H}}$ \cite{zegenhagen1993surface, Bedzyk_phase_structure_factor, bedzyk2004x}, i.e.,
    \begin{equation}\label{F_G}
        F_\textrm{c}^{\boldsymbol{H}} \exp\left(2\pi i P_\textrm{c}^{\boldsymbol{H}} \right) = 
        \mathcal{F}_{\boldsymbol{H}} = 
        \int_{\text{uc}} \rho(\boldsymbol{r}) \exp \left(i {\boldsymbol{H}} \cdot {\boldsymbol{r}}\right) \mathrm{d}\boldsymbol{r}, 
    \end{equation}
    with the integral covering one bulk unit cell (uc), and $\rho(\boldsymbol{r})$ being the distribution function of the atomic species considered. 

    Since $\mathcal{F}_{\boldsymbol{H}}$ also represents the $\boldsymbol{H}^{\text{th}}$-order Fourier coefficient of the distribution function $\rho(\boldsymbol{r})$, eq.\ \eqref{F_G} implies that $\rho(\boldsymbol{r})$ can be reconstructed from the NIXSW measurements by summing up the Fourier components \cite{Bedzyk_phase_structure_factor, bedzyk2004x, Woodruff_2005, ZegenhagenJpn.J.Appl.Phys.2019}:
    \begin{equation}\label{rho}
        \rho(\boldsymbol{r}) = \sum_{\boldsymbol{H}} \mathcal{F}_{\boldsymbol{H}} \exp \left(-i {\boldsymbol{H}} \cdot {\boldsymbol{r}}\right)
                             = \sum_{\boldsymbol{H}} F_\textrm{c}^{\boldsymbol{H}} \exp\left(2\pi i P_\textrm{c}^{\boldsymbol{H}}\right)
                            \exp \left(-i {\boldsymbol{H}} \cdot {\boldsymbol{r}}\right)
                            = 1 + 2\sum_{\substack{\boldsymbol{H} \neq \boldsymbol{0},\\ \boldsymbol{H} \neq -\boldsymbol{H}}}
                            F_\textrm{c}^{\boldsymbol{H}} \cos\left(2\pi P_\textrm{c}^{\boldsymbol{H}} - {\boldsymbol{H}} \cdot {\boldsymbol{r}}\right).
    \end{equation}
    The last simplification introducing the cosine is valid due to Friedel's law, in our case that is $F_\textrm{c}^{\boldsymbol{-H}} = F_\textrm{c}^{\boldsymbol{H}}$ and $P_\textrm{c}^{\boldsymbol{-H}} = -P_\textrm{c}^{\boldsymbol{H}}$ \cite{bedzyk2004x}, and due to $F_\textrm{c}^{\boldsymbol{0}}=$ 1. 
    Of course, only the infinite sum will result in a complete reconstruction of the distribution function, which is experimentally not possible. However, in practice it turns out that -- depending on the complexity of the investigated system -- measurements on a manageable number of less than ten symmetrically inequivalent Bragg reflections are sufficient to solve the structure unambiguously. In the present case of the \textbeta\ and bismuthene phases we have measured on seven independent reflections, the (0004), (10$\overline{\text{1}}$1), ($\overline{\text{1}}$011), (10$\overline{\text{1}}$2), ($\overline{\text{1}}$012), (10$\overline{\text{1}}$3), and ($\overline{\text{1}}$013) reflections. Together with their symmetry equivalent reflections (that are the ($ihkl$) and ($kihl$) reflections), this results in a Fourier sum with 19 components. For all reflections we have recorded NIXSW data for those atomic species discussed above. Corresponding yield curves, Argand diagrams, and a table summarizing all the resulting data are shown in Section 2 of the Supplementary Information.
    
    \begin{figure}[t!]
        \centering
        \includegraphics[scale=1]{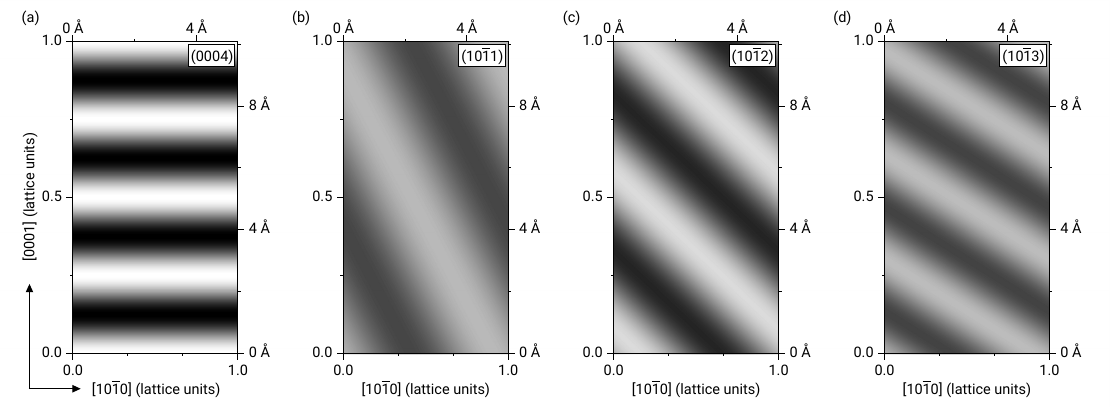}
        \caption{Density distribution maps (2D cuts in the (11$\overline{\text{2}}$0) plane) of selected individual Fourier terms for bulk Si. The distributions correspond to single, one-dimensional cosine functions, the amplitudes and phases of which are given by the coherent fractions and positions obtained for the respective Bragg reflection. The period of the cosine corresponds to the Bragg plane spacing. High (low) densities are shown bright (dark).}
        \label{fig:XSWI_single}
    \end{figure}

    \subsubsection{NIXSW imaging results for the bulk species}

    We start the discussion of the results with the bulk species. 
    \autoref{fig:XSWI_single} illustrates a single component analysis, i.e., the reconstructed atomic density distribution for the case that only one Fourier compound is considered. From eq.\ \eqref{F_G} it is obvious that restricting the sum to $\boldsymbol{H}=$ (0004) results in a cosinusoidal distribution in $z$ direction, see \autoref{fig:XSWI_single}(a), with one maximum lying close to zero on the [0001] axis, since the coherent position (the phase of the cosine term) was measured to be $P_\textrm{c}^{\text{(0004)}}=$ \qty{0.014(2)}{}. The high amplitude of the cosine term reflects the high coherent fraction $F_\textrm{c}^{\text{(0004)}}=$ \qty{1.031(8)}{}. Further maxima are located in distances of multiples of \qty{2.52}{\angstrom}, according to the Bragg plane spacing $d_{\text{(0004)}}$. Hence, this plot illustrates the vertical structure of the bulk Si species with four Si layers in the unit cell. The density distribution maps for the other reflections (\autoref{fig:XSWI_single}(b-d)) show similar scenarios for their respective direction of the $\boldsymbol{H}$ vector and the corresponding Bragg plane spacing $d_{\boldsymbol{H}}$. Note that the smaller amplitudes (smaller coherent fractions) in these three cases indicate that not all Si atoms lie in the same distance to the Bragg planes in these directions. Hence, in contrast to the case of the (0004) reflection, here we cannot directly conclude on atomic positions or distances from the individual plots. However, from these plots it becomes clear that lateral structural information is added by the NIXSW measurements on these inclined reflections having their Bragg planes not parallel to the surface.
    
\begin{figure}[t!]
        \centering
        \includegraphics[scale=1]{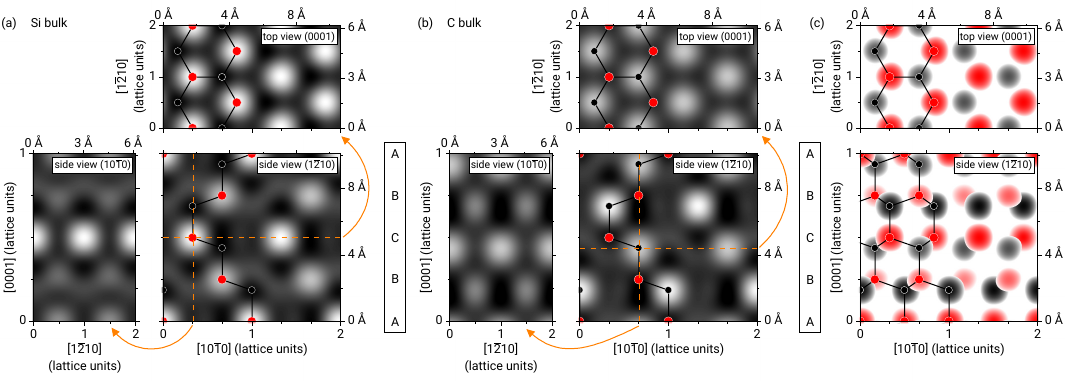}
        \caption{Geometric structure of 4H-SiC as determined by NIXSW imaging. In (a) and (b) selected cross-sectional views of the determined atomic distribution for Si and C, respectively, are shown. The main image (bottom right) displays the (1$\overline{\text{2}}$10) plane, whereas the left and top image show cuts in perpendicular directions as indicated by the orange dashed lines. The superimposed ball-and-stick models demonstrate an excellent agreement between the expected atomic positions and the maxima of the determined atomic distribution. In (c) a combined illustration of the maxima from (a) and (b), color coded as red and black for Si and C, is presented. Note that for the lower image, maxima of an additional plane offset by $a/2$ were included, to resemble the typical side view of 4H-SiC often shown in ball-and-stick models.   }
        \label{fig:XSWI_cuts}
    \end{figure}

    Summing up the four maps of \autoref{fig:XSWI_single}(and the remaining 15 ones that are not shown), that is, extending the Fourier sum to all available summands, results in the lower right plot shown in \autoref{fig:XSWI_cuts}(a) (labeled ``side view (1$\overline{\text{2}}$10)'').
    This map represents a cut through the Fourier-reconstructed density distribution map of the Si bulk species in the (1$\overline{\text{2}}$10) plane that is spanned by the surface normal [0001] and the [10$\overline{\text{1}}$0] directions. In this plane the bulk layers A, B and C are lying horizontally, but are shifted with respect to each other. The other two maps represent perpendicular planes, namely the (0001) plane (top view onto the unit cell, upper map) and another side view, the (10$\overline{\text{1}}$0) plane (left). The dashed orange lines in the (1$\overline{\text{2}}$10) map indicate the intersection lines of the other two planes.

    The interference of the various cosine terms in the Fourier sum now produces clearly separated maxima, which indicate the positions of the bulk Si atoms in three dimensions and reproduce the expected structure of the Si atoms in the 4H-SiC bulk crystal very well, as illustrated by the superimposed ball-and-stick model in the (1$\overline{\text{2}}$10) projection. Also the lateral shift of neighboring layers in [10$\overline{\text{1}}$0] direction according to the ABCBA stacking is reproduced correctly. The same is the case for the bulk C atoms, as can be seen in Fig.\ \ref{fig:XSWI_cuts}(b), which shows the result of the analysis of NIXSW data set recorded for the C~1s bulk component. In the side views the patterns for Si and C are very similar, but shifted vertically according to the Si-C bonding distances in the bulk structure. When comparing the top view maps of (a) and (b), one also finds a shift between the hexagonal patterns of Si and C atoms. One species appears at hollow sites of the other. This is due to the $z$-position of the horizontal cuts, which were chosen within the same bilayer (layer C in this case), as indicated by the dashed orange lines in the (1$\overline{\text{2}}$10) maps. 
    Note that for silicon the maxima of the B layers appear slightly weaker compared to those of the A and C layers, for carbon it is vice versa. This is most likely due to the finite number of Fourier components used in the analysis. 
    
    The full 3D structure of the SiC bulk crystal is obtained when the reconstructed density distributions for both species are combined. In \autoref{fig:XSWI_cuts}(c) we present a top and side view (the (1$\overline{\text{2}}$10) and (10$\overline{\text{1}}$0) planes, respectively), with the distributions of both species superimposed, Si in red and C in black. The intensity scaling is chosen in a way that only clear maxima of the individual density distributions are visible. Furthermore, for the side view not only the density in the (1$\overline{\text{2}}$10) plane is shown, but additionally from a parallel plane with an offset of $a/2$. This produces the characteristic ``3-dimensional'' side view of 4H-SiC, with all atoms within the unit cell being displayed. The varying sizes of the maxima only reflect the intensity differences in the individual density distribution maps reported above.

    \subsubsection{NIXSW imaging results for Bi}

\begin{figure}[t!]
        \centering
        \includegraphics[scale=1]{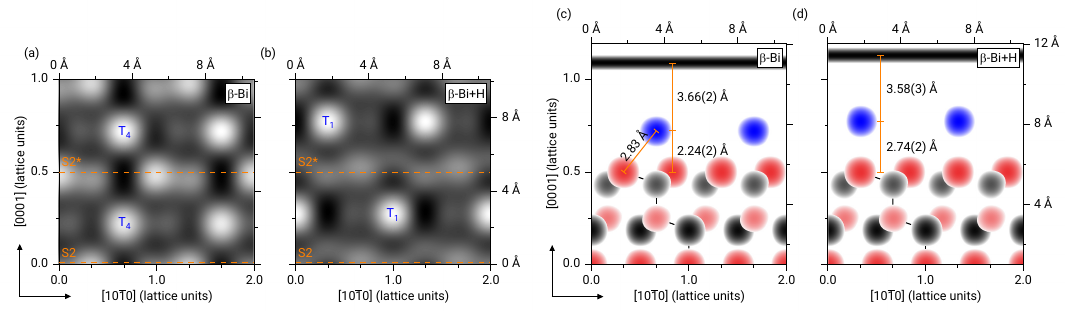}
        \caption{Comparison of the Bi adsorbate structure for both phases, as determined by NIXSW imaging. In (a) and (b) the atomic distributions for Bi are shown for the \textbeta\ phase and bismuthene, respectively (same cross-sectional views as in \autoref{fig:XSWI_cuts}(a) and (b)). S2 and S2* terminations of the bulk are indicated by orange dashed lines. In (c) and (d) the combined bulk and surface structure is shown for a S2* terminated surface, with Si, C, and Bi color coded in red, black, and blue, respectively. The change from the $\text{T}_\text{4}$ to the $\text{T}_\text{1}$ adsorption site due to hydrogenation is clearly visible. For the topmost graphene layer no lateral atomic positions can be determined since it is incommensurate with the substrate.
        }
        \label{fig:XSWI_Bi}
    \end{figure}

    So far we have used the NIXSW data set recorded for the bulk species in order to explain the principle of the NIXSW imaging technique and to demonstrate that the SiC structure can be unambiguously reconstructed by this model-free approach. But of course we also applied the method to determine the atomic sites occupied by Bi. In \autoref{fig:XSWI_Bi}(a) we show the result obtained for Bi, based on NIXSW imaging data recorded from the Bi\,4f$_\text{7/2}$ core levels. As in \autoref{fig:XSWI_cuts}(a), the atomic distribution in the (1$\overline{\text{2}}$10) plane is shown for the \textbeta\ phase prior to hydrogenation. The two strongest maxima that can be seen are located close to the maxima found for the Si bulk species, at coordinates (0.67, 0.22) and (0.67, 0.72) in units of the SiC unit cell, and are labeled ``$\text{T}_\text{4}$''. These positions must be interpreted as follows:  
    
    Since NIXSW imaging is based on the bulk Bragg reflections, it is only able to determine atomic positions within the periodicity of the bulk unit cell, and thus the atomic density obtained by the Fourier-sum analysis is projected into the unit cell.  For a surface species, however, the surface termination of the bulk crystal has to be considered in order to interpret the obtained positions correctly. In our case, the (0001)-oriented 4H-SiC bulk exhibits two dominant surface terminations, namely the so called S2 and S2* terminations, which are characterized by the A and C planes forming the surface plane, respectively. Both terminations are indicated by orange dashed lines in \autoref{fig:XSWI_Bi}(a). The obtained positions of Bi have now to be interpreted with respect to these surface terminations, which makes clear that the lower of the two Bi peaks must be understood as an adsorbate on the S2 terminated surface, while the upper Bi peak indicates an adsorbate on the S2* termination. Comparing these positions of Bi with those of the bulk species reveals that they are horizontally aligned with the Si atoms in the B layers, hence, they correspond to $\text{T}_\text{4}$ hollow sites with a distance of \qty{2.24(2)}{\angstrom} above the bulk-terminating Si plane. 
    Note that there are additional maxima visible in the distribution map, weaker, but still with significant intensity, almost coinciding with the A and C plane. Whether these are merely artefacts resulting from the limited number of Fourier components, or correspond to Bi atoms on $\text{T}_\text{4}$ sites above the (parasitic) S1/S1*-terminated areas cannot be unambiguously clarified.  But the fact that for some of the Bragg reflections, namely the (10$\overline{\text{1}}$1), ($\overline{\text{1}}$011), (10$\overline{\text{1}}$3), and ($\overline{\text{1}}$013) (and their equivalent reflections), the measured coherent fractions were quite small speaks for the former since this reduces the number of Fourier terms with relevant amplitudes to only seven in this case.

    \autoref{fig:XSWI_Bi}(b) illustrates the Bi-based NIXSW imaging results on the bismuthene phase (i.e., after hydrogenation) in a density plot similar to \autoref{fig:XSWI_Bi}(a). Here, the primary peaks are clearly shifted, horizontally by precisely $\pm$ 1/3 of the unit cell, and also slightly upwards. The new positions (0.00, 0.27) and (0.33, 0.77) correspond to on top sites of the terminating Si atoms, i.e., $\text{T}_\text{1}$ adsorption sites. The vertical distance between the peak and the surface plane has increased to \qty{2.74(2)}{\angstrom}.

    In \autoref{fig:XSWI_Bi}(c) and (d) we have combined the NIXSW imaging results for bulk and surface species. For the case of an S2* terminated surface the uppermost SiC bilayers ares shown in a side view in red and black, similar to \autoref{fig:XSWI_cuts}(c). The corresponding Bi adsorption sites are added as blue maxima in their, and the graphene layer is shown as a black bar at the top. For the latter no lateral positions can be determined since it is incommensurate with the substrate. 
    
    These two images clearly illustrate the change of the Bi adsorption site in the phase transition, from the $\text{T}_\text{4}$ site (above the 2$^\text{nd}$-layer Si atoms) in the \textbeta\ phase to the $\text{T}_\text{1}$ site (above the 1$^\text{st}$-layer Si atoms) in bismuthene. This change of the adsorption site is the decisive structural feature enabling the formation of truly 2D bismuthene (see below and Ref.\ \cite{Tilgner2025}). The 3D density maps also allow a straightforward calculation of bonding distances.
    On the $\text{T}_\text{4}$ site, when each Bi atom saturates three Si dangling bonds, we find a Bi-Si distance of \qty{2.83}{\angstrom}. This value is significantly larger than the sum of the covalent radii for Si and Bi, \qty{2.67}{\angstrom} \cite{pyykko2009molecular}. After hydrogenation, when Bi sits on the $\text{T}_\text{1}$ on-top site, we find a stronger Si-Bi bond with its length reduced to \qty{2.73}{\angstrom}, close to the expected covalent bond length.
    
\section{Conclusion}

    The main result of our comprehensive study of the graphene-protected \textbeta\ and bismuthene phases on the (0001) oriented 4H-SiC surface is the determination of the adsorption sites of Bi in both phases. We find that it changes from the $\text{T}_\text{4}$ hollow site in the \textbeta\ phase to the $\text{T}_\text{1}$ on-top site in bismuthene. This change implies a different coordination of Bi with the underlying top-most Si atoms. At the $\text{T}_\text{4}$ hollow site each Bi atoms has three equidistant Si neighbors and thus saturates three dangling bonds. But in the hydrogenation process, Bi is depleted from the hollow site since hydrogen, with its higher electron affinity, saturates one third of all dangling bonds. Bi is thus moving to an on-top site and forms only one bond to the Si atom underneath ($\text{T}_\text{1}$ on-top adsorption site). The remaining four valence electrons of the Bi atoms thus form a graphene-like honeycomb structure which transforms the \textbeta\ phase layer into truly 2D bismuthene. The formation of this energetically very stable layer is most likely the reason that hydrogen is only able to saturate 1/3 of the available Si dangling bonds, and cannot deplete all Bi atoms from the SiC interface. The change of the adsorption site, caused by the hydrogenation of one Si dangling bond per unit cell, is the driving force for the formation of the bismuthene layer. 

    The key finding of this study was obtained from NIXSW imaging, a method that was suggested more than two decades ago, but hardly applied since then. Unlike all standard diffraction methods, can retrieve not only the amplitudes but also the phases of the structure factors \cite{Bedzyk_phase_structure_factor, bedzyk2004x}, and hence allows the precise localization of atomic sites relative to the Bragg planes, independently for all atomic species. The technique is model free, i.e., no model refinement has to be performed, and hence is not prone to false predictions due to local minima in the goodness-of-fit function as it is in principal the case for any structure refinement technique.
    Instead, the element-specific, three-dimensional atomic distribution is obtained, simply as the Fourier-sum of the complex structure factors (amplitudes and phases) that were experimentally obtained for a certain set of Bragg reflections. In general, the reconstruction is the better the more Fourier components can be obtained, and is limited to the bulk unit cell since it is based on bulk Bragg reflections. The latter limitation--in practice--is not a crucial shortcoming, as long as the superstructure unit cell of the investigated surface structure is not too large. Under this condition, NIXSW imaging is a very useful, reliably and unique tool for (surface) structure determination, as we demonstrated here for the prominent example of the QSHI bismuthene.
    
\section*{Methods}	
	
	\subsection*{Sample Preparation}

    4H-SiC wafer pieces were purchased from Pam-Xiamen. The epitaxial graphene synthesis was carried out using the polymer-assisted sublimation growth, which is described in detail elsewhere \cite{Kruskopf2DMater.2016}. Bi was intercalated using an in situ deposition and annealing approach, as reported in previous studies \cite{SohnJ.KoreanPhys.Soc.2021, Wolff_2024}: The deposition was performed in a dedicated evaporation chamber with a base pressure below \qty{5e{-9}}{mbar}, using a custom-built Knudsen cell heated to \qty{550}{\degreeCelsius} for \qty{120}{min}. Subsequent to in-vacuo transfer to the UHV analysis chamber, intercalation of the \textalpha\ phase was executed by heating the sample to \qty{450}{\degreeCelsius} for \qty{30}{min}. The \textbeta\ phase transformation was accomplished by additional annealing to \qty{950}{\degreeCelsius} for \qty{20}{min}. A pyrometer was used to monitor the temperature, assuming a sample emissivity of 0.9. After brief transport through air, hydrogen intercalation was performed using a contactless infrared heating system \cite{diss_sieber} by annealing the sample at \qty{550}{\degreeCelsius} under ultra-pure hydrogen atmosphere (\qty{880}{mbar}, \qty{0.9}{slm}) for \qty{90}{min}.

    \subsection*{Normal Incidence X-ray Standing Wave}

    All experiments presented in this work were performed at the I09 beamline of the Diamond Light Source Ltd. (Didcot, UK) under UHV conditions. The samples were transported to the beamline under UHV conditions after being prepared at the TU Chemnitz. Once the samples were transferred and outgassed, they were positioned in the X-ray beam in a way that they satisfied the Bragg condition for one specific reflection in near-normal-incidence geometry. Note that in order to separate the back-diffracted X-ray beam from the incident X-ray beam, the experiments were conducted slightly off-normal incidence (\qty{3.5}{\degree}). Then, we scanned the photon energy for each reflection within a small energy range of $\pm$\qty{1.5}{eV} around the Bragg energy. For each energy step, the XP spectra of C\,1s, Si\,2s, and Bi\,4f as well as the intensity of the back-diffracted X-ray beam were recorded. A Scienta EW4000 HAXPES hemispherical electron analyzer was used to detect the photoelectrons. Analyzing the XP spectra and extracting the intensities of the relevant core level peaks resulted in one photoelectron yield curve per species. This procedure was repeated for all selected Bragg reflection, resulting in a full NIXSW imaging data set. The samples were maintained at room temperature throughout the entire measurements. The X-ray beam spot size was \qty{400}{\micro\meter} $\times$ \qty{400}{\micro\meter} when incident normal to the sample surface.

\section*{Author contributions}	

    N.T., T.S., P.S., and C.K.\ conceived the project. 
    The sample preparation was carried out by N.T.
    The NIXSW experiments were performed by N.T, S.W., S.S., T.-L.L., and C.K., and analyzed by N.T.\ and S.W., with significant input from S.S.\ and C.K.
    The atomistic model explaining the phase transition was developed by N.T., F.S.T., T.S., P.S., and C.K. 
    All authors discussed the results. N.T.\ and S.W.\ made the figures, and N.T., F.G., P.S., and C.K.\ wrote the paper, with significant input from all authors.

\section*{Acknowledgments}	
	
	The authors thank Christoph Lohse for his contributions to substrate preparation. We also thank the Diamond Light Source for granting access to beamline I09 under proposal SI36085-2 and extend our appreciation to the I09 beamline staff (Pardeep Kumar Thakur and Dave McCue) for their valuable support. This work was supported by the German Research Foundation (Deutsche Forschungsgemeinschaft, DFG) within the Research Unit FOR5242 (project 449119662) and the Collaborative Research Centre SFB-1083 (project A12).

\begin{singlespace}

\begin{thebibliography}{10}
	\providecommand{\url}[1]{\texttt{#1}}
	\providecommand{\urlprefix}{URL }
	
	\bibitem{arnal2019ab}
	R.~D. Arnal, M.~Metz, A.~J. Morgan, H.~N. Chapman, and R.~P. Millane: Ab initio
	phasing using diffraction data from different crystal forms. \emph{2019
		International Conference on Image and Vision Computing New Zealand (IVCNZ)},
	1--6, IEEE (2019).
	
	\bibitem{fienup1982phase}
	J.~R. Fienup: Phase retrieval algorithms: a comparison. \emph{Applied optics}
	\textbf{21}(15), 2758 (1982).
	
	\bibitem{elser2003phase}
	V.~Elser: Phase retrieval by iterated projections. \emph{JOSA A}
	\textbf{20}(1), 40 (2003).
	
	\bibitem{oszlanyi2004ab}
	G.~Oszl{\'a}nyi and A.~S{\"u}t{\H{o}}: Ab initio structure solution by charge
	flipping. \emph{Acta Crystallographica Section A: Foundations of
		Crystallography} \textbf{60}(2), 134 (2004).
	
	\bibitem{patterson1934fourier}
	A.~L. Patterson: A Fourier series method for the determination of the
	components of interatomic distances in crystals. \emph{Physical Review}
	\textbf{46}(5), 372 (1934).
	
	\bibitem{batterman1969detection}
	B.~W. Batterman: Detection of foreign atom sites by their x-ray fluorescence
	scattering. \emph{Physical review letters} \textbf{22}(14), 703 (1969).
	
	\bibitem{zegenhagen1993surface}
	J.~Zegenhagen: Surface structure determination with X-ray standing waves.
	\emph{Surface Science Reports} \textbf{18}(7-8), 202 (1993).
	
	\bibitem{vartanyants2001theory}
	I.~Vartanyants and M.~Kovalchuk: Theory and applications of x-ray standing
	waves in real crystals. \emph{Reports on Progress in Physics} \textbf{64}(9),
	1009 (2001).
	
	\bibitem{Woodruff_2005}
	D.~P. Woodruff: Surface structure determination using x-ray standing waves.
	\emph{Reports on Progress in Physics} \textbf{68}(4), 743 (2005),
	\urlprefix\url{https://dx.doi.org/10.1088/0034-4885/68/4/R01}.
	
	\bibitem{Zegenhagen2013}
	J.~Zegenhagen and A.~Kazimirov: \emph{{X-Ray Standing Wave Technique :
			Principles and Applications}}. World Scientific Publishing Company (2013).
	
	\bibitem{LinPhys.Rev.B2022}
	Y.-R. Lin, S.~Wolff, P.~Schädlich, M.~Hutter, S.~Soubatch, T.-L. Lee, F.~S.
	Tautz, T.~Seyller, C.~Kumpf, and F.~C. Bocquet: Vertical structure of
	Sb-intercalated quasifreestanding graphene on {SiC}(0001). \emph{Phys. Rev.
		B} \textbf{106}(15), 155418 (2022).
	
	\bibitem{SchaedlichAdv.Mater.Interfaces2023}
	P.~Schädlich, C.~Ghosal, M.~Stettner, B.~Matta, S.~Wolff, F.~Schölzel,
	P.~Richter, M.~Hutter, A.~Haags, S.~Wenzel, Z.~Mamiyev, J.~Koch, S.~Soubatch,
	P.~Rosenzweig, C.~Polley, F.~S. Tautz, C.~Kumpf, K.~Küster, U.~Starke,
	T.~Seyller, F.~C. Bocquet, and C.~Tegenkamp: Domain Boundary Formation Within
	an Intercalated {Pb} Monolayer Featuring Charge-Neutral Epitaxial Graphene.
	\emph{Adv. Mater. Interfaces}  (2023).
	
	\bibitem{Wolff_2024}
	S.~Wolff, M.~Hutter, P.~Schädlich, H.~Yin, M.~Stettner, S.~Wenzel, F.~S.
	Tautz, F.~C. Bocquet, T.~Seyller, and C.~Kumpf: Bi-intercalated epitaxial
	graphene on SiC(0001). \emph{New Journal of Physics} \textbf{26}(10), 103009
	(2024), \urlprefix\url{https://dx.doi.org/10.1088/1367-2630/ad7f7d}.
	
	\bibitem{grossmann2022evolution}
	L.~Grossmann, D.~A. Duncan, S.~P. Jarvis, R.~G. Jones, S.~De, J.~Rosen,
	M.~Schmittel, W.~M. Heckl, J.~Bj{\"o}rk, and M.~Lackinger: Evolution of
	adsorption heights in the on-surface synthesis and decoupling of covalent
	organic networks on Ag (111) by normal-incidence X-ray standing wave.
	\emph{Nanoscale horizons} \textbf{7}(1), 51 (2022).
	
	\bibitem{ding2023does}
	P.~Ding, M.~Braim, A.~Hobson, L.~Rochford, P.~Ryan, D.~Duncan, T.-L. Lee,
	H.~Hussain, G.~Costantini, M.~Yu, \emph{et~al.}: Does F4TCNQ Adsorption on Cu
	(111) Form a 2D-MOF? \emph{The Journal of Physical Chemistry C}
	\textbf{127}(42), 20903 (2023).
	
	\bibitem{hBN_on_Ni}
	M.~Raths, C.~Schott, J.~Knippertz, M.~Franke, Y.-R. Lin, A.~Haags,
	M.~Aeschlimann, C.~Kumpf, and B.~Stadtmüller: Growth, domain structure, and
	atomic adsorption sites of hBN on the Ni(111) surface. \emph{Physical Review
		Materials} \textbf{5} (2021).
	
	\bibitem{XSW_of_Si_on_GaAs}
	M.~Sugiyama, S.~Maeyama, and M.~Oshima: {X‐ray standing wave study of a
		Si‐adsorbed GaAs(001) surface}. \emph{Applied Physics Letters}
	\textbf{68}(26), 3731 (1996),
	\urlprefix\url{https://doi.org/10.1063/1.115988}.
	
	\bibitem{woicik1994structural}
	J.~Woicik, G.~Franklin, C.~Liu, R.~Martinez, I.-S. Hwong, M.~Bedzyk, J.~Patel,
	and J.~A. Golovchenko: Structural determination of the Si (111)
	($\sqrt{3}\times\sqrt{3}$)-Bi surface by x-ray standing waves and scanning tunneling
	microscopy. \emph{Physical Review B} \textbf{50}(16), 12246 (1994).
	
	\bibitem{bedzyk2004x}
	M.~Bedzyk, P.~Fenter, Z.~Zhang, L.~Cheng, J.~Okasinski, and N.~Sturchio: X-ray
	Standing Wave Imaging. \emph{Synchrotron Radiation News} \textbf{17}(3), 5
	(2004), \urlprefix\url{https://doi.org/10.1080/08940880408603088}.
	
	\bibitem{Bedzyk_phase_structure_factor}
	M.~J. Bedzyk and G.~Materlik: Two-beam dynamical diffraction solution of the
	phase problem: A determination with x-ray standing-wave fields. \emph{Phys.
		Rev. B} \textbf{32}, 6456 (1985),
	\urlprefix\url{https://link.aps.org/doi/10.1103/PhysRevB.32.6456}.
	
	\bibitem{SohnJ.KoreanPhys.Soc.2021}
	Y.~Sohn, S.~W. Jung, F.~Göhler, W.~J. Shin, S.~Cha, T.~Seyller, and K.~S. Kim:
	Electronic band structure of {Bi}-intercalate layers in graphene and
	{SiC}(0001). \emph{J. Korean Phys. Soc.} \textbf{78}(2), 157 (2021).
	
	\bibitem{RiedlPhys.Rev.Lett.2009}
	C.~Riedl, C.~Coletti, T.~Iwasaki, A.~A. Zakharov, and U.~Starke:
	Quasi-Free-Standing Epitaxial Graphene on {SiC} Obtained by Hydrogen
	Intercalation. \emph{Phys. Rev. Lett.} \textbf{103}(24), 246804 (2009).
	
	\bibitem{BriggsNat.Mater.2020}
	N.~Briggs, B.~Bersch, Y.~Wang, J.~Jiang, R.~J. Koch, N.~Nayir, K.~Wang,
	M.~Kolmer, W.~Ko, A.~D. L.~F. Duran, S.~Subramanian, C.~Dong,
	J.~Shallenberger, M.~Fu, Q.~Zou, Y.-W. Chuang, Z.~Gai, A.-P. Li, A.~Bostwick,
	C.~Jozwiak, C.-Z. Chang, E.~Rotenberg, J.~Zhu, A.~C.~T. van Duin, V.~Crespi,
	and J.~A. Robinson: Atomically thin half-van der Waals metals enabled by
	confinement heteroepitaxy. \emph{Nat. Mater.} \textbf{19}(6), 637 (2020).
	
	\bibitem{EmtsevNat.Mater.2009}
	K.~V. Emtsev, A.~Bostwick, K.~Horn, J.~Jobst, G.~L. Kellogg, L.~Ley, J.~L.
	McChesney, T.~Ohta, S.~A. Reshanov, J.~R\"ohrl, E.~Rotenberg, A.~K. Schmid,
	D.~Waldmann, H.~B. Weber, and T.~Seyller: Towards wafer-size graphene layers
	by atmospheric pressure graphitization of silicon carbide. \emph{Nat. Mater.}
	\textbf{8}(3), 203 (2009).
	
	\bibitem{ReisScience2017}
	F.~Reis, G.~Li, L.~Dudy, M.~Bauernfeind, S.~Glass, W.~Hanke, R.~Thomale,
	J.~Schäfer, and R.~Claessen: Bismuthene on a SiC substrate: A candidate for
	a high-temperature quantum spin Hall material. \emph{Science}
	\textbf{357}(6348), 287 (2017).
	
	\bibitem{Tilgner2025}
	N.~Tilgner, S.~Wolff, S.~Soubatch, T.-L. Lee, A.~D.~P. Unigarro, S.~Gemming,
	F.~S. Tautz, C.~Kumpf, T.~Seyller, F.~Göhler, and P.~Schädlich: Reversible
	Switching of the Environment-Protected Quantum Spin Hall Insulator Bismuthene
	at the Graphene/SiC Interface  (2025).
	
	\bibitem{pyykko2009molecular}
	P.~Pyykk{\"o} and M.~Atsumi: Molecular single-bond covalent radii for elements
	1--118. \emph{Chemistry--A European Journal} \textbf{15}(1), 186 (2009).
	
	\bibitem{bondi1964van}
	A.~v. Bondi: van der Waals Volumes and Radii. \emph{The Journal of physical
		chemistry} \textbf{68}(3), 441 (1964).
	
	\bibitem{mantina2009consistent}
	M.~Mantina, A.~C. Chamberlin, R.~Valero, C.~J. Cramer, and D.~G. Truhlar:
	Consistent van der Waals radii for the whole main group. \emph{The Journal of
		Physical Chemistry A} \textbf{113}(19), 5806 (2009).
	
	\bibitem{SforziniPhys.Rev.Lett.2015}
	J.~Sforzini, L.~Nemec, T.~Denig, B.~Stadtmüller, T.-L. Lee, C.~Kumpf,
	S.~Soubatch, U.~Starke, P.~Rinke, V.~Blum, F.~Bocquet, and F.~Tautz:
	Approaching Truly Freestanding Graphene: The Structure of
	Hydrogen-Intercalated Graphene on {6H}-{SiC}(0001). \emph{Phys. Rev. Lett.}
	\textbf{114}(10), 106804 (2015).
	
	\bibitem{SforziniPhys.Rev.Lett.2016}
	J.~Sforzini, P.~Hapala, M.~Franke, G.~van Straaten, A.~St\"ohr, S.~Link,
	S.~Soubatch, P.~Jel\'{\i}nek, T.-L. Lee, U.~Starke,
	M.~\ifmmode~\check{S}\else \v{S}\fi{}vec, F.~C. Bocquet, and F.~S. Tautz:
	Structural and Electronic Properties of Nitrogen-Doped Graphene. \emph{Phys.
		Rev. Lett.} \textbf{116}(12), 126805 (2016).
	
	\bibitem{ZegenhagenJpn.J.Appl.Phys.2019}
	J.~Zegenhagen: X-ray standing waves technique: Fourier imaging active sites.
	\emph{Jpn. J. Appl. Phys.} \textbf{58}(11), 110502 (2019).
	
	\bibitem{Kruskopf2DMater.2016}
	M.~Kruskopf, D.~M. Pakdehi, K.~Pierz, S.~Wundrack, R.~Stosch, T.~Dziomba,
	M.~Götz, J.~Baringhaus, J.~Aprojanz, C.~Tegenkamp, J.~Lidzba, T.~Seyller,
	F.~Hohls, F.~J. Ahlers, and H.~W. Schumacher: Comeback of epitaxial graphene
	for electronics: large-area growth of bilayer-free graphene on {SiC}.
	\emph{2D Mater.} \textbf{3}(4), 041002 (2016).
	
	\bibitem{diss_sieber}
	N.~Sieber: \emph{Wasserstoff- und Sauerstoffstabilisierte 6H-SiC(0001)
		Oberflächen - Eine Studie Chemischer, Struktureller und Elektronischer
		Eigenschaften}. Dissertation, Friedrich-Alexander-Universit\"at
	Erlangen-N\"urnberg (2002).
	
\end{thebibliography}

\end{singlespace}

\end{document}


\maketitle

\section{Fitting models for the core-level photoelectron spectra}

    As discussed in the main text, annealing of the sample required for the \textalpha\ to \textbeta\ phase transformation also triggers a partial deintercalation~\cite{Wolff_2024} resulting in localized domains of the zeroth layer graphene (ZLG). Correspondingly, the C\,1s fitting models of the \textbeta\ phase before hydrogenation (Figure 1 a) include the following components: 
    $\text{C}^{\text{ZLG}}$,
    $\text{C}^{\text{QFG}}_\text{Bi}$, 
    $\text{C}^{\text{bulk}}_{\text{ZLG}}$, and
    $\text{C}^{\text{bulk}}_\text{Bi}$
    representing, respectively, the carbon of ZLG and Bi-intercalated QFG, and the carbon of the bulk underneath the domains of ZLG and Bi-intercalated QFG. 
    The Si\,2s core level (Figure 1 b) of the bulk requires two components 
    $\text{Si}^{\text{bulk}}_\text{Bi}$ and
    $\text{Si}^{\text{bulk}}_{\text{ZLG}}$,
    while for Bi\,4f$_\text{7/2}$ a single component is sufficient.
    Upon subsequent hydrogenation, the ZLG domains convert to hydrogen intercalated quasi-freestanding graphene (QFG). Thus, the  $\text{C}^{\text{ZLG}}$ component disappears and instead the new $\text{C}^{\text{QFG}}_\text{H}$ component develops in the C\,1s spectrum. Also the $\text{C}^{\text{bulk}}_{\text{ZLG}}$ vanishes, but the corresponding bulk-C component underneath H-intercalated graphene cannot be distinguished from its counterpart underneath Bi, and hence only one component, $\text{C}^{\text{bulk}}$, is used for fitting both QFG domains. The same is valid for the Si core levels. 
       
    In the fitting model, the contribution of $\text{C}^{\text{QFG}}_\text{Bi}$  assumed to have an asymmetric line shape due to its quasi-metallic nature, was simulated by two symmetric components (solid and dashed blue lines in Figure 1 a,c) separated by 0.3 eV and with a fixed intensity ratio.  This mitigates parasitic contributions from the ZLG to the asymmetric tail of the graphene signal.     
    The energy difference between the two bulk components $\text{C}^{\text{bulk}}_\text{Bi}$ and $\text{C}^{\text{bulk}}_{\text{ZLG}}$ was constrained to \qty{0.8}{eV}, according to different surface band bendings for the bulk beneath the intercalated and ZLG areas \cite{RisteinPhys.Rev.Lett.2012}. The corresponding Si bulk core levels were fitted in an analogous way. For the Bi signal, a single symmetric profile was used, representing the insulating Bi layer of the \textbeta\ phase \cite{Wolff_2024, SohnJ.KoreanPhys.Soc.2021}.
    For the bismuthene phase, the position of the $\text{C}^{\text{QFG}}_\text{H}$ peak was fixed to  \qty{0.4}{eV} below that of the 
    $\text{C}^{\text{QFG}}_\text{Bi}$. This accounts for different doping levels of graphene intercalated by H and Bi, as determined from the relative shift of the  \textpi-bands seen in angle-resolved photoelectron spectroscopy \cite{Tilgner2025}. Regarding the peak asymmetry of the two peaks, the same fitting procedure as discussed above was applied.

\newpage

\section{NIXSW results for inclined Bragg reflections}

    \begin{figure}[h!]
        \centering
        \includegraphics[scale=1]{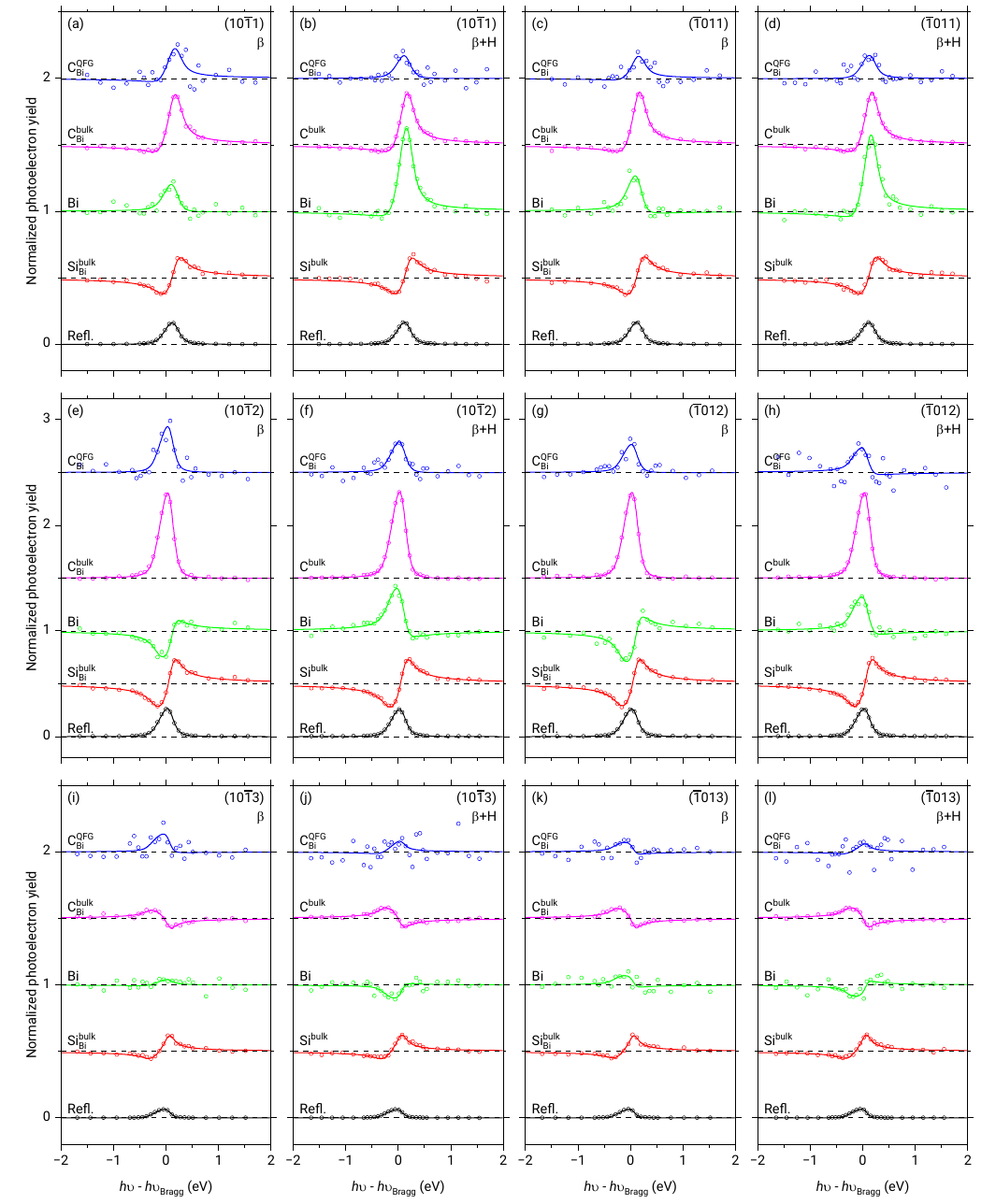}
        \caption{Representative reflectivities and photoelectron yield curves for the six measured inclined reflections and the relevant species. (a, c, e, g, i, k) display the results for the \textbeta\ phase, whereas (b, d, f, h, j, l) show the results for bismuthene (\textbeta +H). The Bragg energy for (10$\overline{\text{1}}$1) and ($\overline{\text{1}}$011) is \qty{2.4081}{keV}, for (10$\overline{\text{1}}$2) and ($\overline{\text{1}}$012) \qty{2.6338}{keV}, and for (10$\overline{\text{1}}$3) and ($\overline{\text{1}}$013) \qty{2.9721}{keV}.}
        \label{fig:full_yield_refl}
    \end{figure}

    \newpage

    \begin{landscape}
        \begin{figure}[t!]
        \centering
        \includegraphics[scale=1]{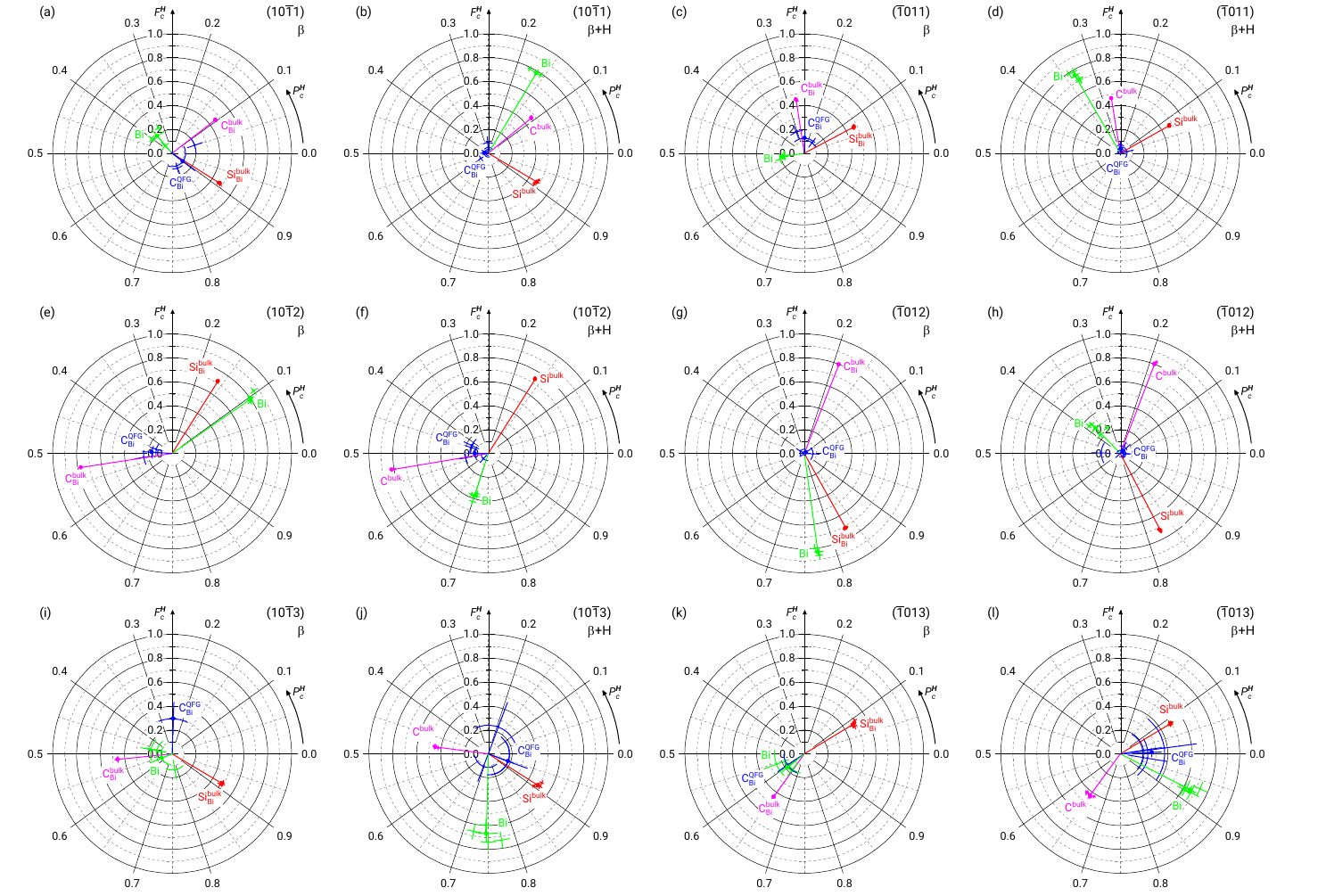}
        \caption{Argand diagrams for the six measured inclined reflections and the relevant species. (a, c, e, g, i, k) display the results for the \textbeta\ phase, whereas (b, d, f, h, j, l) show the results for bismuthene (\textbeta +H). The plots show the results for each individual measurement (dots) as well as the averaged values (vectors) used for the NIXSW imaging.}
        \label{fig:full_argand}
    \end{figure}
    \end{landscape}

    \newpage

    We performed normal incidence X-ray standing wave (NIXSW) experiments of the \textbeta\ phase and bismuthene using the reflections  (0004), (10$\overline{\text{1}}$1), ($\overline{\text{1}}$011), (10$\overline{\text{1}}$2), ($\overline{\text{1}}$012), (10$\overline{\text{1}}$3), and ($\overline{\text{1}}$013). The first reflection, (0004), is perpendicular to the sample surface, while the others are inclined.
    Note that in the redundant coordinate system of the hexagonal lattice, the reflections ($hkil$), ($ihkl$), and ($kihl$) are equivalent~\cite{StarkePhys.StatusSolidiB1997}. Hence, in addition to $\mathcal{F}_{\text{(0004)}}$, we acquired 18 Fourier components by using the six inequivalent inclined reflections.
    Corresponding reflectivities and photoelectron yield curves are shown in \autoref{fig:full_yield_refl}, while the respective complex structure factors are visualized in the Argand diagrams in \autoref{fig:full_argand}. 
    The mean values of coherent fractions and coherent positions averaged over several measurements for different species and reflections are listed in \autoref{tab_beta} and \autoref{tab_beta_H} for the \textbeta\ phase and bismuthene, respectively. These are the values used for NIXSW imaging. Note that only the photoelectron yield of Bi is noticeably affected by hydrogenation (cf.\ green curves in~\autoref{fig:full_yield_refl} for the same reflections before and after hydrogenation, e.g., \autoref{fig:full_yield_refl} (a) and (b) for the (10$\overline{\text{1}}$1) reflection) pointing at the phase transformation in the Bi layer, while the bulk components remain unaffected. 
    Specifically, NIXSW analysis reveals that the coherent positions of Bi change drastically (\autoref{fig:full_argand}) suggesting a change of adsorption site. Any effect of hydrogenation on $\text{C}^{\text{QFG}}_\text{Bi}$ observed in Argand diagrams is barely meaningful due to low coherent fractions and great scatter of the corresponding results attributed to the incoherence of the graphene sheet with respect to the bulk lattice for the included reflections. 
    
    \begin{table}[h!]
        \centering
        \caption{Averaged coherent fractions and positions for all measured reflections on the \textbeta\ phase before hydrogenation, as obtained from several NIXSW measurements.}
        \label{tab_beta}
        \renewcommand{\arraystretch}{1.2} 
        \begin{tabular}{|p{1cm}|p{1.5cm}|p{1.5cm}|p{1.5cm}|p{1.5cm}|p{1.5cm}|p{1.5cm}|p{1.5cm}|p{1.5cm}|} 
            \hline
             & \multicolumn{2}{c|}{$\text{Si}_{\text{Bi}}^\text{bulk}$} & \multicolumn{2}{c|}{$\text{C}_{\text{Bi}}^\text{bulk}$} & \multicolumn{2}{c|}{Bi} & \multicolumn{2}{c|}{$\text{C}_{\text{Bi}}^\text{QFG}$} \\ 
            \hline
                      & $F_\textrm{c}^{\boldsymbol{H}}$ & $P_\textrm{c}^{\boldsymbol{H}}$ & $F_\textrm{c}^{\boldsymbol{H}}$ & $P_\textrm{c}^{\boldsymbol{H}}$ & $F_\textrm{c}^{\boldsymbol{H}}$ & $P_\textrm{c}^{\boldsymbol{H}}$ & $F_\textrm{c}^{\boldsymbol{H}}$ & $P_\textrm{c}^{\boldsymbol{H}}$ \\
            \hline
            (0004)  & 1.031(8)  & 0.014(2)  & 0.908(5)  & 0.758(1)  & 1.025(12)  & 0.902(3)  & 0.98(6)  & 0.355(8)  \\ 
            \hline
            (10$\overline{\text{1}}$1)  & 0.470(12)  & 0.910(2) & 0.459(1) & 0.105(3) & 0.18(7) & 0.37(3) & 0.17(4) & 0.93(13) \\ 
            \hline
            ($\overline{\text{1}}$011)  & 0.469(7) & 0.077(3) & 0.452(10) & 0.274(2) & 0.183(19) & 0.525(10) & 0.15(4) & 0.23(8) \\ 
            \hline
            (10$\overline{\text{1}}$2)  & 0.715(6) & 0.160(1) & 0.772(4) & 0.525(1) & 0.81(4) & 0.098(5) & 0.19(6) & 0.49(4) \\ 
            \hline
            ($\overline{\text{1}}$012)  & 0.714(3) & 0.830(3) & 0.799(1) & 0.192(1) & 0.83(3) & 0.772(3) & 0.06(2) & 0.2(3) \\ 
            \hline
            (10$\overline{\text{1}}$3)  & 0.486(5) & 0.914(4) & 0.46(3) & 0.52(1) & 0.15(3) & 0.57(15) & 0.30(14) & 0.24(7) \\ 
            \hline
            ($\overline{\text{1}}$013)  & 0.484(17) & 0.086(6) & 0.442(16) & 0.652(7) & 0.20(4) & 0.60(4) & 0.17(10) & 0.59(10) \\ 
            \hline
        \end{tabular}
    \end{table}

    \begin{table}[h!]
        \centering
        \caption{Averaged coherent fractions and positions for all measured reflections on the bismuthene phase (after hydrogenation), as obtained from several NIXSW measurements.}
        \label{tab_beta_H}
        \renewcommand{\arraystretch}{1.2} 
        \begin{tabular}{|p{1cm}|p{1.5cm}|p{1.5cm}|p{1.5cm}|p{1.5cm}|p{1.5cm}|p{1.5cm}|p{1.5cm}|p{1.5cm}|} 
            \hline
             & \multicolumn{2}{c|}{$\text{Si}^\text{bulk}$} & \multicolumn{2}{c|}{$\text{C}^\text{bulk}$} & \multicolumn{2}{c|}{Bi} & \multicolumn{2}{c|}{$\text{C}_{\text{Bi}}^\text{QFG}$} \\ 
            \hline
                      & $F_\textrm{c}^{\boldsymbol{H}}$ & $P_\textrm{c}^{\boldsymbol{H}}$ & $F_\textrm{c}^{\boldsymbol{H}}$ & $P_\textrm{c}^{\boldsymbol{H}}$ & $F_\textrm{c}^{\boldsymbol{H}}$ & $P_\textrm{c}^{\boldsymbol{H}}$ & $F_\textrm{c}^{\boldsymbol{H}}$ & $P_\textrm{c}^{\boldsymbol{H}}$ \\
            \hline
            (0004)  & 1.040(7)  & 0.006(3)  & 0.947(9)  & 0.757(1)  & 1.159(8)  & 0.093(1)  & 0.84(5)  & 0.512(3)  \\ 
            \hline
            (10$\overline{\text{1}}$1)  & 0.471(7)  & 0.913(6) & 0.466(6) & 0.109(3) & 0.785(3) & 0.163(9) & 0.06(3) & 0.39(15) \\ 
            \hline
            ($\overline{\text{1}}$011)  & 0.471(6) & 0.082(2) & 0.464(4) & 0.276(1) & 0.75(4) & 0.333(5) & 0.04(3) & 0.20(10) \\ 
            \hline
            (10$\overline{\text{1}}$2)  & 0.736(7) & 0.161(1) & 0.817(15) & 0.527(1) & 0.38(3) & 0.701(8) & 0.14(5) & 0.51(9) \\ 
            \hline
            ($\overline{\text{1}}$012)  & 0.724(10) & 0.826(1) & 0.806(15) & 0.191(1) & 0.29(6) & 0.37(3) & 0.10(3) & 0.2(3) \\ 
            \hline
            (10$\overline{\text{1}}$3)  & 0.494(17) & 0.911(7) & 0.446(13) & 0.480(2) & 0.68(6) & 0.747(19) & 0.18(6) & 0.9(3) \\ 
            \hline
            ($\overline{\text{1}}$013)  & 0.492(8) & 0.087(2) & 0.430(9) & 0.648(11) & 0.66(5) & 0.924(6) & 0.27(8) & 0.00(3) \\ 
            \hline
        \end{tabular}
    \end{table}

    A set of Bragg planes is defined by a reciprocal lattice vector $\boldsymbol{H} = h\boldsymbol{b}_1 + k\boldsymbol{b}_2 + l\boldsymbol{b}_3$ with the reciprocal unit cell vectors $\boldsymbol{b}_i$. For the hexagonal unit cell of 4H-SiC and relative to a cartesian coordinate system they are given by 
    \begin{equation}\label{rez_vec}
        \boldsymbol{b}_1 = \frac{2\pi}{\sqrt{3}a}
        \left(\begin{smallmatrix}
            \sqrt{3} \\
            -1 \\
            0
        \end{smallmatrix}\right),\,\,
        \boldsymbol{b}_2 = \frac{2\pi}{\sqrt{3}a}
        \left(\begin{smallmatrix}
            \sqrt{3} \\
            1 \\
            0
        \end{smallmatrix}\right),\,\,
        \text{and}\,\,\,\,\boldsymbol{b}_3 = \frac{2\pi}{c}
        \left(\begin{smallmatrix}
            0 \\
            0 \\
            1
        \end{smallmatrix}\right),
    \end{equation}
    where $a = \text{\qty{3.08}{\angstrom}}$ and $c = \text{\qty{10.08}{\angstrom}}$ are the lattice constants \cite{Peng_Lou_Jin_Wang_Wang_Wang_Chen_2009}. Note that $|\boldsymbol{H}| = 2\pi / d_{\boldsymbol{H}}$ with the Bragg plane spacing $d_{\boldsymbol{H}}$.

\newpage

\section{Effect of the 4H-SiC surface termination}

    \begin{figure}[h!]
        \centering
        \includegraphics[scale=1]{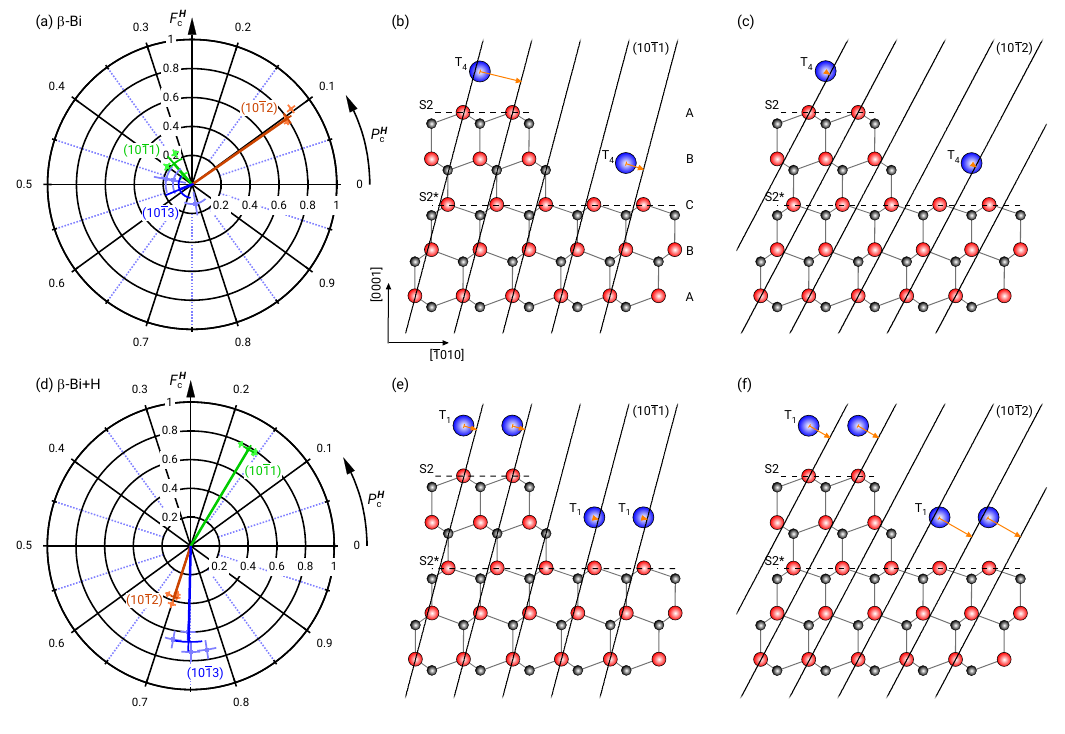}
        \caption{ Effect of two possible terminations of 4H-SiC surface on NIXSW results of Bi. (a) Argand diagram of Bi \textbeta\ phase for three inclined reflections and (b,c) relative position of Bi atoms at the T$_\text{4}$ adsorption site with respect to (b) (10$\overline{\text{1}}$1) and (c) (10$\overline{\text{1}}$2) Bragg planes. 
        (d) Argand diagram of bismuthene and (e,f) relative position of Bi atoms at the T$_\text{1}$ adsorption site with respect to (e) (10$\overline{\text{1}}$1) and (f) (10$\overline{\text{1}}$2) Bragg planes. 
        The (10$\overline{\text{1}}$1) and (10$\overline{\text{1}}$2) Bragg planes intersect the crystal perpendicular to the figure plane.}
        \label{fig:planes}
    \end{figure}

    \autoref{fig:planes} illustrates the effect of two possible terminations S2 and S2* of 4H-SiC crystal~\cite{ramsdell1947studies, PakdehiAdv.Funct.Mater.2020} on the coherent fraction of Bi before and after hydrogenation. 
    For the \textbeta\ phase (\autoref{fig:planes} a-c), Bi atoms at the T$_\text{4}$ adsorption sites on two differently terminated surfaces are located at significantly different distances from the (10$\overline{\text{1}}$1) Bragg plane (see red arrows at the Bi atoms labeled T$_\text{4}$) resulting in a very low coherent fraction. Contrary, with respect to the (10$\overline{\text{1}}$2) Bragg plane, their heights are identical, and, accordingly,  the coherent fraction is very high. 
    After hydrogenation (\autoref{fig:planes} d-f) resulting in the displacement of Bi from T$_\text{4}$ to T$_\text{1}$ adsorption site, for two terminations Bi is located at  different distances with respect to the (10$\overline{\text{1}}$2) plane, while its height over the (10$\overline{\text{1}}$1) plane is only marginally different. 
    Finally, for NIXSW with inclined Bragg reflections, the transformation of the Bi layer triggered by hydrogenation 
    manifests itself  not only in the coherent positions but also quite strongly in the coherent fractions. 
    Note that if the 4H-SiC surface had only one termination, either S2 or S2*, the effect of hydrogenation on the coherent fractions of Bi would not be so large but primarily reflecting the structural quality of the phase. The change of adsorption sites, however, would still be very well visible in different coherent positions.

\newpage

\begin{singlespace}

\end{singlespace}